\documentclass[prb,twocolumn,floatfix,showpacs]{revtex4}
\usepackage{graphicx,natbib}
\begin{document}

\title{Chiral Mixed Phase in Disordered $3d$ Heisenberg Models}

\author{S. Bekhechi}
\author{ B.W. Southern}
\affiliation{Department of Physics and Astronomy, University of Manitoba,
Winnipeg Manitoba, Canada R3T 2N2}
\date{\today}
\begin{abstract}
 Using Monte Carlo simulations, we compute the spin stiffness of a site-random 3d Heisenberg model with
competing ferromagnetic and antiferromagnetic interactions. Our results for the pure limit yield values of the 
  the critical temperature and the critical exponent $\nu$  in excellent  agreement with previous high precision
studies. In the disordered case, a mixed "chiral" phase is found which may be in the same universality class
as $3d$ Heisenberg spin glasses.

\end{abstract}


\maketitle


The critical behavior of magnetic systems with quenched disorder has been of considerable theoretical
and experimental interest for over 30 years. Disorder can lead to a competition between two order parameters
and to a phase diagram which has regions in which each one orders independently as well as a mixed phase where both
order simultaneously. These phases often meet at multicritical points and can have various properties which depend on
the spin dimensionality. Aharony \cite{aharony03} has recently reviewed some of the old and new results on
multicritical points with special attention to high-$T_c$ materials. However, the role played by quenched randomness
remains a open question. 

Early renormalization group studies of systems  with isotropic interactions and
quenched disorder indicated that  competing phases existed with multicritical points described by complex exponents\cite{chen77} or the absence of a stable fixed point\cite{aharony80}. Various scenarios were suggested which included first order transitions, smeared transitions or spin glass ordering. For many years it was believed that a spin glass phase does not exist in $3d$ XY and
Heisenberg systems. However, numerical studies\cite{kawa00} during the last decade have suggested that
chiral degrees of freedom are important in spin glasses which have continuous degrees of freedom. 
More recent work\cite{leeyoung03,endoh01} indicates that  Heisenberg systems exhibit a finite temperature
spin glass transition in three dimensions where both the spin and chiral degrees of freedom order simultaneously. The correlation
length critical exponent  is estimated to be $\nu = 1.1(2)$ which differs substantially from that of the pure $3d$ Heisenberg ferromagnet.

  In the present work we report a  finite size scaling study of the spin
  stiffness of a $3d$ site random isotropic Heisenberg model introduced previously by Matsubara {\it et. al.}\cite{matsu96}. The model  describes a mixture of $A$
   and $B$ magnetic ions randomly distributed at the sites of a simple cubic lattice with concentrations $p$ and $1-p$ respectively. The exchange
   bonds between neighboring ion pairs are defined so that $+J$, $-J$, and $+J$  correspond to  $A-A$, $B-B$,
   and $A-B$ (or $B-A$)  pairs respectively. The Hamitonian can be written as
\begin{equation}
H = -\frac{J}{2} \sum_{<i,j>} [1+(\varepsilon_{i}+\varepsilon_{j}) - \varepsilon_{i}\varepsilon_{j}] \vec{S_{i}}\cdot \vec{S_{j}} 
\end{equation}
where  $\vec{S}_i=(S^{x}_i,S^{y}_i,S^{z}_i)$ represents a classical three component spin of unit magnitude 
located at each site $i$ and $\varepsilon_{i}=1$ or $-1$ for $A$ or $B$ ions respectively. The average value of $\varepsilon_i$
over the lattice is $<\varepsilon_i>=2p-1$. As discussed by the previous authors, the Hamiltonian has a symmetry with respect
to $p$ and $(1-p)$ and hence the phase diagram is symmetric about $p=0.5$ as shown schematically in Fig. 1 of their paper. There are three ordered phases corresponding to
conventional ferromagnetism, antiferromagnetism and a mixed intermediate phase in which both a longitudinal ferromagnetism and transverse antiferromagnetism exist.
The previous work used a finite size scaling analysis
of the order parameter cumulants to determine the critical behavior at these transitions.

We 
  show that the spin stiffness for vector spin models can provide an independent alternative to the usual 
order parameter cumulants for locating the critical 
  temperature  and determining the critical exponent $\nu$ associated with the correlation length.
Our results for the phase diagram and critical exponent $\nu$ differ from those reported previously.

  We divide the bipartite lattice into two identical interpenetrating
 sublattices $\alpha$ and $\beta$ and consider the following order parameters
\begin{eqnarray}
 \vec{m}=\frac{1}{N}(\sum_{i\in \alpha} \vec{S_i} + \sum_{i\in \beta} \vec{S_i}) \nonumber \\
\vec{m}_s=\frac{1}{N}(\sum_{i\in \alpha} \vec{S_i} - \sum_{i\in \beta} \vec{S_i})
\end{eqnarray}
where $\vec{m}$ is the total magnetization per site and $\vec{m}_s$ is the staggered magnetization per site. These two order parameters describe the pure non-disordered phases at $p=1,0$ respectively.  Due to the symmetry about $p=0.5$ we shall only consider the region $p \geq 0.5$.

 The spin stiffness (helicity) tensor can be written\cite{SY93} as an equilibrium correlation function by taking
 second derivatives of the free energy with respect to the strength of an imposed spin gradient involving a twist  about a
 particular direction in spin space. We choose an orthogonal set of spin axes which correspond to unit vectors in the directions of 
$\vec{m}_s \times \vec{m}$,  $\vec{m}\times (\vec{m}_s\times \vec{m})$ and $\vec{m}$ which we denote as $\hat{1}, 
\hat{2}$  and $\hat{3}$ respectively. Hence $\hat{2}$ is a unit vector in the direction of the component of the staggered
magnetization which is perpendicular to $\vec{m}$. The spin stiffness elements can be written as

\begin{widetext}
 \begin{eqnarray} 
 \rho_{a b}&=& \frac{1}{L^{3}} 
 \sum_{i<j} J_{ij} (\hat{e}_{ij}\cdot \hat{u})^{2}
 <\vec{S_{i}} \cdot \vec{S_{j}}\  \delta_{a, b} 
 - \frac{S_{i}^{a} S_{i}^{b}+S_{i}^{b} S_{j}^{a}}{2}> \nonumber \\
   &  & 
\mbox{} - \frac{1}{T L^{3}} <\sum_{i<j} J_{ij}
 (\hat{e}_{ij}\cdot \hat{u})
 (\vec{S_{i}}\times \vec{S_{j}})^{a} 
   \mbox{} \sum_{l<m} J_{lm} (\hat{e}_{lm}\cdot
 \hat{u})(\vec{S_{l}} \times \vec{S_{m}})^{b}>
 \end{eqnarray}
\end{widetext}
 where $J_{ij}=\frac{J}{2}[1+(\varepsilon_{i}+\varepsilon_{j}) - \varepsilon_{i}\varepsilon_{j}]$ and the angular brackets $<...>$ indicate
a thermal average. The superscripts $a,b$ denote the components of the spin vectors in the orthonormal system 
 $\hat{1}, \hat{2}, \hat{3}$. The $\hat{e}_{ij}$ are unit vectors along neighboring bonds and $\hat{u}$ is the
 spatial direction of the gradient in the lattice.

\begin{figure}[b]
\centering
\includegraphics[height=65mm,width=85mm]{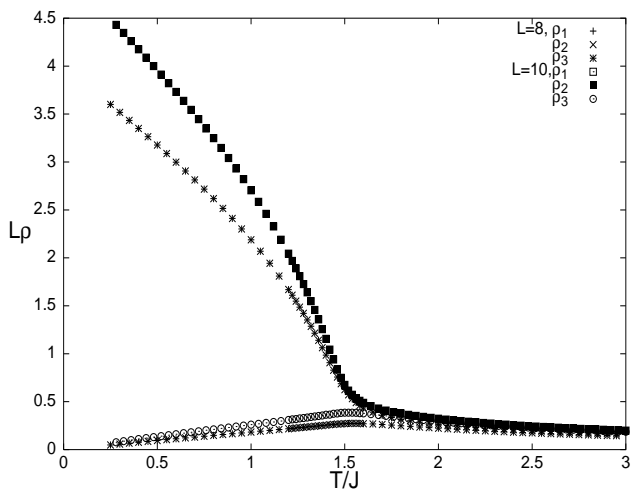}
\caption{The transverse and longitudinal spin stiffnesses multiplied by $L$ for $L=8$ and $L=10$ at $p=1$ as a function of $T/J$. }
\end{figure}

\begin{figure}[b]
\centering
\includegraphics[height=65mm,width=85mm]{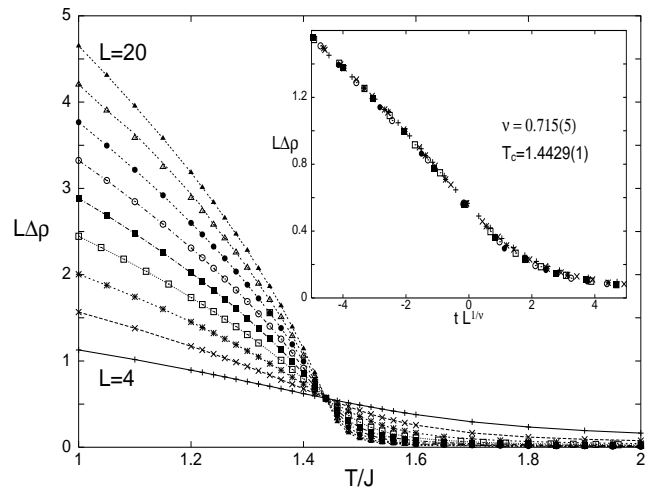}
\caption{The difference between transverse and longitudinal stiffnesses times $L$ as a function of $T/J$ for sizes $L=4, 6, 8, ... 20$. The insert shows a finite size scaling collapse
 of the stiffnesses  onto a single curve for $T_c=1.4429(1)$ and $\nu=0.715(5)$. }
\end{figure}  

{\it Pure Heisenberg Model}

We first consider the pure limit $p=1$ which describes the $3d$ ferromagnetic Heisenberg model. In this
case there is no staggered magnetization and the two directions perpendicular  to $\vec{m}$ are equivalent.
We find that the spin stiffness tensor is diagonal with the principal values $\rho_{1}=\rho_{2} \neq \rho_3$.
 The three principal stiffnesses are computed for simple cubic lattices with periodic boundary conditions and
linear  dimensions $L$ ranging from 6 to 32 using a single-spin flip heat bath algorithm\cite{olive}. We used $1-5\times
  10^5$ Monte Carlo steps (MCS) for performing our measurements after discarding the first
  $5\times10^4$ MCS to reach thermal equilibrium. In order to have continuous
   functions of temperature we
  also applied  histogram reweighting techniques \cite{n2f}  near $T_c$ using longer runs of $(2-5 \times 10^7)$ MCS in order 
   to improve the statistics. Since we have continuous degrees of freedom, the histograms
   are built up by binning the values of the stiffnesses.

 Finite size scaling considerations for the stiffness $\rho(T,L)$
 predict that the singular part behaves as
 \begin{equation}
 \rho(T,L)=\frac{1}{L}f(L^{1/\nu}t),
 \end{equation}
 where $t$ is the reduced temperature and $\nu$ is the correlation length exponent. This scaling form predicts that plots  of  $L \rho(T,L)$ versus $T$ should intersect at $T_c$ and that the $L$ dependence of the slopes of the curves at this crossing point can be used to determine $\nu$.  Fig. 1 shows the three principal stiffnesses multiplied by $L$ as a function of $T$ for sizes $L=8, 10$. 
The longitudinal stiffness
$L \rho_3$ exhibits a small peak which saturates with larger sizes. Both transverse stiffnesses behave identically and exhibit size dependent increases at low $T$ proportional to $L$. Using the scaling form in (4) for $L \rho_1$ and $L \rho_2$ does not lead to a crossing point for different $L$.
However, if we interpret the behavior of the longitudinal component $L \rho_3$ as a regular term and use the same scaling form for
the difference $\Delta\rho$ between the transverse and longitudinal components, we obtain the results shown in Fig 2.  The results clearly show
 a crossing temperature which moves to higher temperature as $L$ is increased.


  Assuming that $L \Delta\rho(L,T)$ has the same phenomenological behavior as the Binder cumulant
  \cite{n2g}, we can determine the crossing point for different sizes $L,L'$ and extrapolate $T_c(L)$ to infinite size
 in terms of  the quantity $(\ln L'/L)^{-1}$ where $L$ is the smaller of the two sizes. For $L=8,10$ and $12$ we
find a  good linear behavior for $T_c(L)$.
Using this method, the critical temperature is estimated to be
  $T_{c}=1.4429(1)$ and this value is in good agreement with the 
  results from  high-resolution Monte Carlo studies\cite{n3a,n3b}. 
  The derivative of $L\Delta\rho$ with respect to $t$ should behave as  $L\Delta\rho'\approx L^{1/\nu}f'(0)$ and
this prediction is obeyed quite well. 
The value of the static exponent $\nu$
 is estimated to be $\nu=0.715(5)$ which is remarkably close to the recently
  improved theoretical value\cite{n2a}. Using these values of $T_c$ and $\nu$, 
  we obtain a nice collapse of the data of Fig. 2 into a single universal curve as shown in the inset.

  
{\it Disordered Model} 
 
\begin{figure}[b]
\centering
\includegraphics[height=65mm,width=85mm]{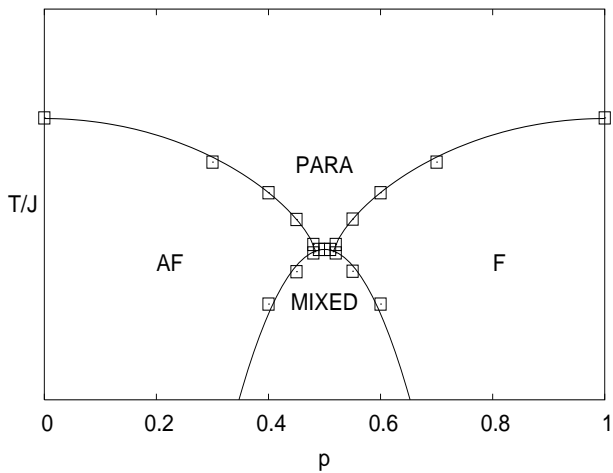}
\caption{Schematic phase diagram of the site disordered Heisenberg system.}
\end{figure}

The exponents obtained above for the pure Heisenberg model yield a negative value of $\alpha=2-d\nu \sim -.15$. The Harris
criterion\cite{harris74} predicts that the critical behavior of the paramagnetic-ferromagnetic transition in the disordered system should have
the same values if the specific heat exponent of the pure system is negative. Using the method described above, we have studied lattice sizes ranging from $L=4$
to $L=16$ by averaging the principal values of the stiffness tensor over 500 (200) disorder configurations for small (large) sizes in the range $0.5 \leq p <1$.  For $p=.7$  the finite size scaling analysis described above yields the values
$T_c = 1.217(4), \nu = .70(1)$ and again allows for a nice collapse of the data with the pure Heisenberg exponents. This behavior is in agreement with the predictions of the Harris criterion.

Fig. 3 shows a schematic phase diagram for the model which differs from that reported by Matsubara {\it et. al.}\cite{matsu96}. We find both ordered and mixed phases but they do not meet at a decoupled tetracritical point\cite{aharony03}. 
At values of $p$ slightly larger than $0.5$, two transitions are observed. The upper transition $T_c$ is from the paramagnetic phase to a
ferromagnetic phase where the transverse stiffnesses become different from the longitudinal stiffness. At a lower temperature,
the two transverse components also acquire a difference which signals an order-order transition to a mixed phase.
As indicated in Fig. 3, the upper transition terminates close to $p=0.5$ and there is a finite range of $p$
where there is again a single transition to a mixed phase. We have studied the spin-spin structure factor of this phase and it has two sharp peaks corresponding to the  two order parameters in equation (2). These order parameters are orthogonal to one another and hence the phase is
characterized by a global chirality $\vec{m}_s \times \vec{m}$ and corresponds to a noncollinear order parameter. 


\begin{figure}[t]
\centering
\includegraphics[height=63mm,width=80mm]{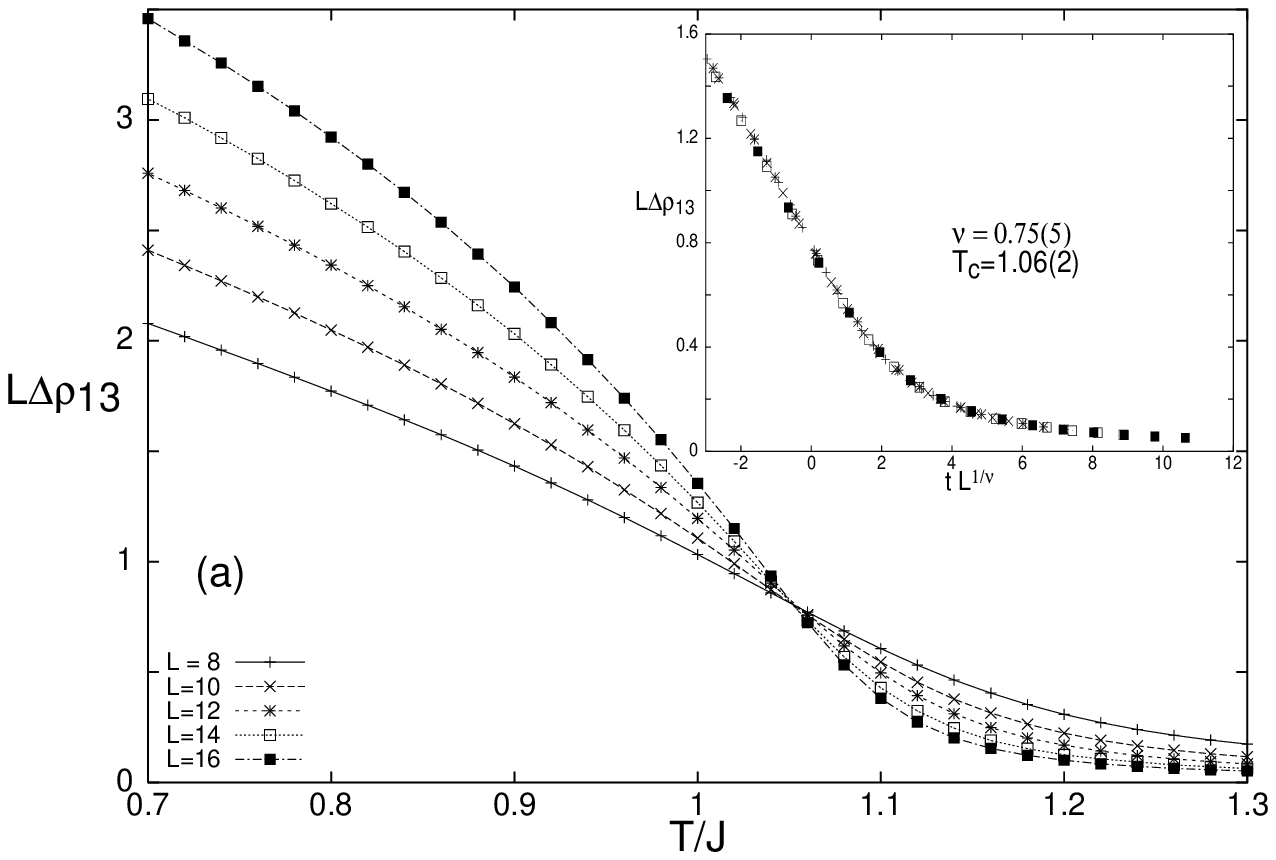}
\includegraphics[height=63mm,width=80mm]{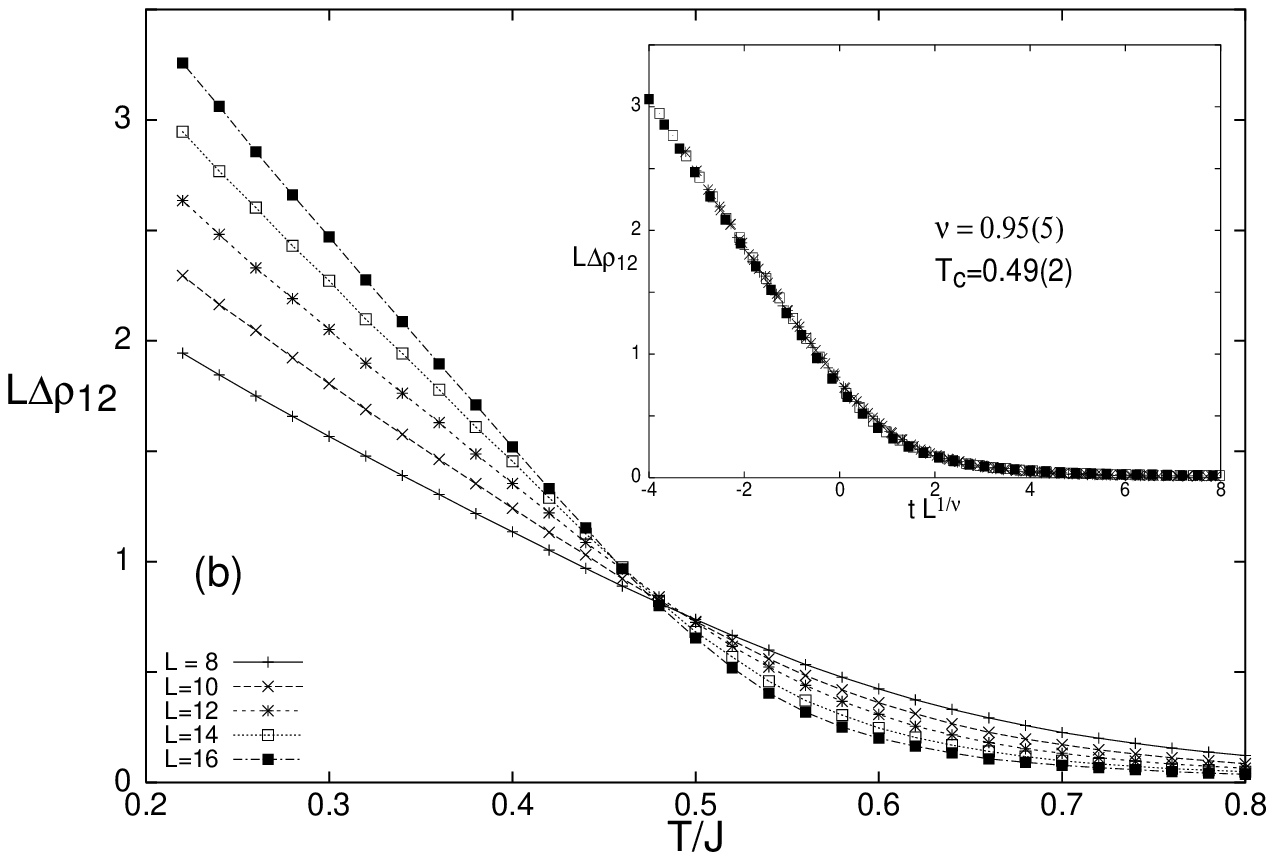}
\caption{$L \Delta \rho$ plotted as a function of $T/J$ for sizes $L$ ranging from $8$ to $16$ for $p=0.6$. The insets show a finite size
scaling collapse of the data.  (a) the upper transition has $T_c=1.06(2)$ and $\nu=0.75(5)$   (b) the lower transition has $T_N=0.49(2)$ and $\nu=0.95(5)$.}
\end{figure}

The stiffness results for $p=0.6$ are shown in Figs. $4 (a), (b)$, where 
it is clearly seen that  two  transitions are detected from the crossing of the stiffnesses $\Delta \rho_{13}$ at $T_c \sim 1.06$ and 
$\Delta \rho_{12}$ at $T_N \sim 0.49$. In Fig. 4 (a) $\Delta \rho_{13}$ is the difference between the transverse and longitudinal
stiffnesses whereas in Fig. 4 (b) $\Delta \rho_{12}$ is the difference between the two transverse stiffnesses.
These values of the critical temperatures agree with those reported previously\cite{matsu96} for this same model at the equivalent concentration $p=0.4$ but our
value of the
critical exponent $\nu$ at the upper transition is quite different. The previous work reported a value of $\nu = 0.54$ which corresponds to
a divergent specific heat. We find no such divergence and our value of $\nu = 0.75(5)$ is compatible with the exponent of the
pure ferromagnet as expected from the Harris criterion. Our value of $\nu = 0.95(5)$ at the lower transition is somewhat larger than that reported by the  previous authors.   Hence the transition to the mixed "chiral" phase is characterized by a value
of $\nu \sim 1$ which is close to the value reported for Heisenberg spin glasses and  may belong to the same universality class.

\begin{table}[t]
\caption{Results for the  critical temperatures $T_c$, $T_N$ and the 
 exponents $\nu$ and $\beta$ for various values of  $p$. }
  \begin{ruledtabular}
  \begin{tabular}{|c|c|c|c|c|}
\hline
$p$ &$T_c$&$T_N$&$\nu$&$\beta$ \\
\hline 
1& $1.4429(1)$ &$ \sim $ &$0.715(5)$ &$0.366(6)$\\
\hline
0.7& $1.217(4)$ &$ \sim $ &$0.70(1)$ &$0.36(1)$\\
\hline 
0.6& $1.06(2)$ &$ \sim $ &$0.75(5)$ &$0.33(3)$\\
& $ \sim $ &$0.49(2)$ &$0.95(5)$ &$0.38(3)$\\
\hline 
0.55& $0.925(5)$ &$ \sim $ &$0.95(5)$ &$0.36(2)$\\
& $ \sim $ &$0.66(1)$ &$0.99(4)$ &$0.34(3)$\\
\hline 
0.52& $0.795(5)$ &$ \sim $ &$1.03(5)$ &$0.34(3)$\\
 & $ \sim $ &$0.752(5)$ &$0.97(4)$ &$0.36(3)$\\
\hline
0.51&$ 0.767(5)$& $ \sim  $& $0.98(2)$& $0.35(2)$ \\
&$  \sim  $& $ 0.771(6) $& $0.97(4)$& $0.35(3)$ \\
\hline 
0.50& $0.770(3)$ &$0.770(3)$ &$0.96(5)$ &$0.37(2)$\\
\hline 
\hline
 \end{tabular}
 \end{ruledtabular}
 \end{table}

 The values of the critical parameters obtained in the range $0.5 \le p \le 1$ using the stiffness are given in Table I. The critical
exponent $\beta$ is obtained from a finite size scaling of the order parameters in equation (2) and its value  is fairly independent of  $p$ and is close to the pure
Heisenberg value.  The transition to the mixed "chiral" phase at $p=0.5$ where there is only a single transition is also described by a value of $\nu \sim 1$. Using $\nu$ and $\beta$, we estimate  the value of the anomalous dimension exponent $\eta \sim -0.3$ for all transitions to the mixed "chiral" phase. These values of $\nu$ and $\eta$ are similar to those found in many spin glass models\cite{nakamura02,leeyoung03} and
may indicate that they are universal. Although the model studied here has quenched site disorder, it may provide a useful reference
model for the random bond models. By introducing dilution as  well as competing interactions, we do not expect the correlation length exponent to change immediately since the mixed phase has a large negative value of
$\alpha \sim -1$. However, for strong dilution, the order parameter could be much more complicated and the Harris argument
may not apply.

Our use of the helicity modulus in eqn(3) as a measure of the stiffness of the system assumes that the derivatives of the
free energy are evaluated at zero twist which corresponds to the equilibrium state in this model. For spin glasses\cite{reger91} and vortex glasses\cite{olsson03}, the equilibrium state does not necessarily correspond to zero twist and the helicity modulus would average to zero over the distribution of random bonds. The fluctuations of the helicity moduli would have to be studied in this case.

In summary, we have studied both the pure and site disordered $3d$ Heisenberg model on a simple cubic geometry using
Monte Carlo methods. We have shown that the spin stiffness provides a convenient method to determine transition
temperatures as well as the correlation length exponent $\nu$. A mixed chiral phase which exhibits a universal value
of $\nu \sim 1$ is found at intermediate concentrations. Our results differ from those reported previously for the same model.

\begin{acknowledgments}

This work was supported by the Natural Sciences and Research Council of Canada, the University of Manitoba Research
Grants Program 
and the High Performance Computing facilities at the University of Manitoba and HPCVL
Canada. We thank Peter Young for useful discussions.
\end{acknowledgments}

\bibliography{smaine5}

\end{document}